# Peak Electricity Demand and Global Warming in the Industrial and Residential areas of Pune : An Extreme Value Approach

Ayush Maheshwari, Dr. Kamal Kumar Murari, Dr. T. Jayaraman

## Abstract

Industrial and residential activities respond distinctly to electricity demand on temperature. Due to increasing temperature trend on account of global warming, its impact on peak electricity demand is a proxy for effective management of electricity infrastructure. Few studies explore the relationship between electricity demand and temperature changes in industrial areas in India mainly due to the limitation of data. The precise role of industrial and residential activities response to the temperature is not explored in sub-tropical humid climate of India. Here, we show the temperature sensitivity of industrial and residential areas in the city of Pune, Maharashtra by keeping other influencing variables on electricity demand as constant. The study seeks to estimate the behaviour of peak electricity demand with the apparent temperature (AT) using the Extreme Value Theory. Our analysis shows that industrial activities are not much influenced by the temperature whereas residential activities show around 1.5-2% change in average electricity demand with 1 degree rise in AT. Further, we show that peak electricity demand in residential areas, performed using stationary and non-stationary GEV models, are significantly influenced by the rise in temperature. The study shows that with the improvement in data collection, better planning for the future development, accounting for the climate change effects, will enhance the effectiveness of electricity distribution system.  The study is limited to the geographical area of Pune. However, the methods are useful in estimating the peak power load attributed to climate change to other geographical regions located in subtropical and humid climate.

# 1. Introduction

Globally extreme heat days are rising and this is expected to be even worse in the future (Alexander et al. 2006; Coumou and Rahmstorf 2012). This is particularly serious in urban areas where hot days and nights have risen significantly for the period 1973-2013 (Mishra et al., 2015). The temperature extremes have significant impacts on human health (McMichael, 2012) and infrastructure (Handmer et al., 2012). Extreme heat conditions combined with high humidity results in acute health problems such as heat cramps, exhaustion, heat stroke and dehydration (Hondula et al., 2014). Temperature extremes also have an impact on urban infrastructure, particularly electricity supply network (Van et al., 2012). Extreme temperature days will result in more cooling demand at residential areas resulting into increase in consumption of electricity (Li et al., 2012). An understanding of the impact of extreme temperature on electricity infrastructure is therefore a necessary step to address the aspects of vulnerability of electricity infrastructure that are associated with climatic conditions. This is particularly important from the perspective of future climate which is expected to be hotter in many areas.

Existing literature highlights non-linear relationship between temperature and electricity demand. The non-linearity in relation is due to cooling demand during hot conditions and heating demand during cold conditions. Generally, temperature-electricity demand curve is an asymmetric U-shaped curve, which is empirically established for number of urban areas (Wangpattarapong et al. 2008, Valor et al. 2001). Generally, such curve is observed in areas where heating demand is significant. Studies examining temperature-electricity relationships, conducted in colder, warm and mild countries has shown an increase in electricity demand by 0.54%, 1.7% and 0.51% respectively for one degree rise in temperature (Cian et al. 2007). Although this is applicable for most areas, however, the sensitivity of electricity network depends on the penetration of air-conditioning equipment, indoor comfort temperature, thermal quality of building stock. Recognising the fact that most developing countries are located in warmer climates , the demand sensitivity to temperature is expected to be higher with the rise in affluence (Gupta 2012).

Few studies have estimated electricity demand in conditions of extreme temperature or conditions of heat and cold wave. Some of them (Colombo et al. 1999, Segal et al. 1992, Franco and Sanstad 2008, Yabe 2005, Parkpoom et al. 2008, Akbari et al. 1992) have highlighted the relationship of extreme events of temperature with the peak electricity load. Colombo (1999) analysed the frequency of extreme heat and electricity demand in nine cities of Canada using moving and overlapping windows for defining extreme events of temperature. Their study concluded that 3° C increase in daily maximum temperature would lead to a 7% increase in the standard deviation of current peak electricity demand during summer. Moral-Carcedo (Moral-Carcedo et al. 2015) analysed the demand sensitivity of firms with temperature in Spain. The paper concludes that almost all firms are insensitive to low temperatures while higher temperatures increase demand for cooling for 44% of the firms. The paper argues that besides service sector, water collection and treatment plants, food products and beverage manufacturing units, construction and real estate activities, and motor vehicle and rubber manufacturing units are significantly sensitive to the higher temperature. The aggregate results are in line with the literature on industrial insensitivity to temperature.

In India, both observed and climate data points to increased extreme temperature days in the future (Murari et al. 2015,; Revadekar et al., 2012). Days with extreme temperature results in an increase in peak electricity consumption that might lead to increased stress on the electricity infrastructure and thereby the power supply facility. For instance, in July 2012, the massive power outage left almost half of India without electricity. The steep rise in peak electricity demand due to hot weather conditions was considered to be one of the reasons for the grid failure which eventually resulted in disruption of the power supply (Romero 2012). Although, there are many factors such as; household size, income, equipment ownership, that are responsible for affecting electricity demand. Exposure to extreme temperatures is expected to put an additional pressure that may result in mismanagement of services or interruption of services due to failure of infrastructure (CEC 2004).

Although, there are limited studies that provide evidence of impact of temperature rise on electricity demand. Gupta (2014) used a semi-parametric variable coefficient approach to address the impact of temperature rise on electricity demand in Delhi. It assumes a non-parametric temperature model and variables such as income, prices following a linear model. Such combination reduces computational complexity and effective modeling of the temperature-electricity relationship. Gupta (2014) suggested the shifting of U-shaped electricity curve leftwards over the years. It implies that people are now comfortable at lesser temperatures than earlier which can be partly attributed to the ownership of air-coolers/conditioners and increasing purchasing power parity. One of the limitations of this study is the aggregation of electricity demand data which may differ across sectors and it may be underestimating the residential electricity demand.
Increase in peak power demand obligates electric utilities to increase power supply by building additional power plants or increasing efficiency. It is a serious concern due to high cost of marginal power generation. Several studies have analysed the relationship between peak electricity demand and ambient temperatures for Thailand (Parkpoom et al 2008), USA (Akbari et al. 1992), Israel (Segal 1992) and California (Franco and Sanstad 2007). In Thailand (Parkpoom et al 2008), per degree increase in temperature raise peak demand by 4.6% while in USA states such as Washington, Los Angeles observes an increase by 3.3-3.6% (Akbari et al. 1992). Therefore, it is imperative to analyse the relationship between peak demand and temperature apart from the overall relationship.

Studies aggregating the demand data at a city scale skews the demand from residential and industrial activities. It presents a vague understanding of the temperature sensitivities of electricity demand from residential and industrial activities. Metropolitan cities such as Delhi, Mumbai and Pune consists of a large number of industries. Demand analysis at the city level will underestimate the residential sensitivity to temperature. Although, industrial activities such as manufacturing plants, industrial processes are insensitive to rise in temperature. It is imperative to perform a sectoral analysis of temperature sensitivities from industries and residencies to understand a clear picture of demand sensitivity to temperature from different activities. The unique contribution of the paper is to present a sectoral analysis of a metropolitan city for industrial and residential activities.

The current paper analyses the relationship between electricity demand (ED) and apparent temperature (AT), using disaggregated data at feeder level (or division level) for Pune city, which is one of the major industrial city close to Western coast of India. The paper further establishes a case for quantifying the change in peak electricity load due to rise in temperature attributed to global warming, using non-stationary extreme value theory. This work is the first study to present a disaggregated ED for industrial and residential activities at a city level located in a subtropical climate in a developing country.

## 2. Description of study area

In this study, we analyse extreme temperature and peak electricity demand relationship. Pune, situated close to Western coast, is the 7$^{th}$ most populated city in India covering an area of 458 km$^2$. The urban agglomeration consists of the population of about 3.3 million, is considered to be fastest growing urban center in India Singhal et al (2013) (Singhal, Shaleen, Stanley McGreal, and Jim Berry. "Application of a hierarchical model for city competitiveness in cities of India." *Cities* 31 (2013): 114-122. ). The city has a strong base of manufacturing and engineering facilities, currently has about 98 % coverage of electricity with no load shedding. Due to current prospects of growth of the city, it is a part of smart city project of Government of India, one of the implications of it is that the service provider of the city are expected to supply 24*7 electricity supply without any interruption. The climate of Pune falls under moderate climatic zone and experiences hot semi-arid conditions (Govardhan and Gadbail, 2014). April is the hottest month with an average daily maximum temperature of 37.4 °C and a corresponding relative humidity of 19%. Winter months (November, December, January, February) are comfortable during the day and cold at night and do not require institutional heating (Govardhan and Gadbail, 2014).

Electricity demand in Pune has a mix of high industrial, commercial and residential activity. Maharashtra State Electricity Distribution Company (MAHADISCOM), which is a public sector company, is the electricity service provider to the city. In the last five years electricity demand in the city has grown by 23% due to rapid urbanisation, population growth and industrial growth. Energy demand from the residential sector has increased by 36% in the same period[1]. Therefore, MAHADISCOM has developed a zero load-shedding model for Pune urban to provide uninterrupted supply to the city. The electricity distribution in the city is managed on a zonal basis each zone is demarcated by the percentage of feeders which cater to a mix of industrial and residential demands. Further details of zones and their demand characteristics are discussed in the section 3.

## 3. Data and Methods

### 3.1 Data

#### 3.1.1 Climatic data

The daily temperature data and relative humidity was obtained for the Pune meteorological station from the period from January 2008 - December 2012. The data on climatic factors is obtained from *www.TuTiempo.net* which gives station-wise data for all major weather stations in India. The data is collected for the Poona meteorological station located at 18°53' latitude and 73°85' longitude and at an altitude of 559 meters. The climatic data is available without any missing points for the study period. The climatic data is extracted for the daily average temperature, relative humidity, wind speed (at an elevation of 10 m) and water vapor pressure.

---

[1] Analysis of Power Situation in Pune and suggestions for Sustainable Urban Electricity Development
http://mnsblueprint.org/pdf/Solutions%20for%20Electricity%20for%20Pune%20city.pdf

### 3.1.2 Electricity Data

We used feeder wise electricity for Pune zone, available on an hourly basis from January 2008 to December 2012. This dataset is taken from Maharashtra state distribution company (MAHADISCOM) website (http://www.mahadiscom.in/). According to current arrangement of electricity distribution of MAHADISCOM, Pune zone is is divided into three circles namely, Ganeshkhind(Urban), Pune(Rural), Rashtrapeth (Urban) circles. Ganeshkhind circle covers the Pune city as well as industrial areas on the periphery. The Ganeshkhind circle is further divided into four divisions namely, Bhosari, Kothrud, Pimpri and Shivaji Nagar. According to Mahadiscom norms, each feeder is classified as scheduled and unscheduled load shedding. The feeders that supply power to industrial/commercial areas are classified as scheduled load shedding (sheddable) feeders and feeders supplying power to residential feeders are classified under unscheduled load shedding (non-sheddable) feeders (MAHADISCOM, 2012). It was observed in the analysis that characteristics of the feeders are changing during the study period. The feeders which were commercial earlier are changed to residential type and vice-versa. It prevents us from assuming feeders as unit of analysis. To overcome this shortcoming, we adopt division as our analysis unit which is a fairer representation of the area.

We have analysed each division based on the number of feeders showing sheddable and non-sheddable characteristics. The non-sheddable feeders supply to the industries, IT spaces, hotels and retail spaces. These areas belong to industrial and retail activities where electricity load shedding isn't unplanned. On the other hand, residential areas are supplied by the sheddable feeders where unscheduled load shedding happens.

Besides classifying the divisions on feeder characteristics, google maps and ground truthing was used to confirm their residential / industrial nature. Feeder sample data was classified twice during the months of June and December for each year of the study period. Figure 1 shows the graph for the number of feeders with sheddable and non-sheddable characteristics.

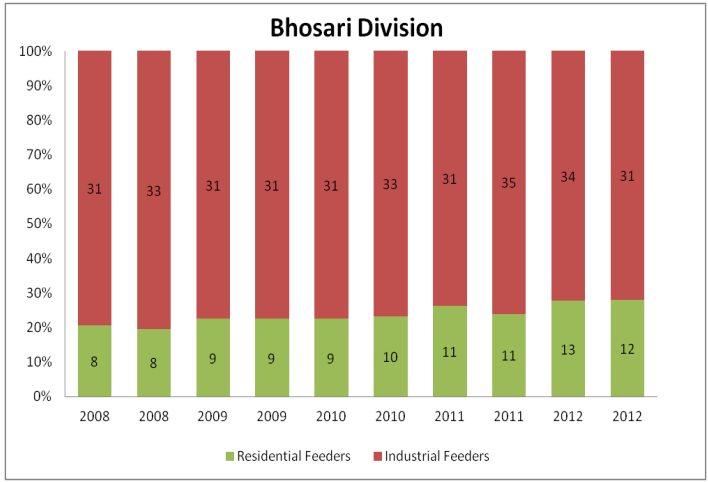
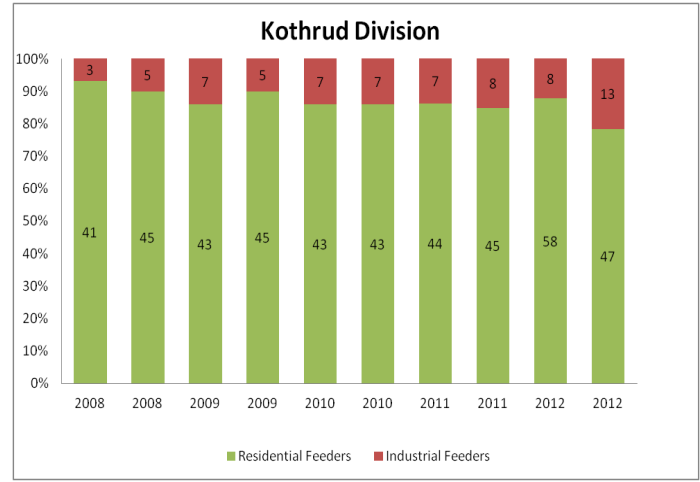
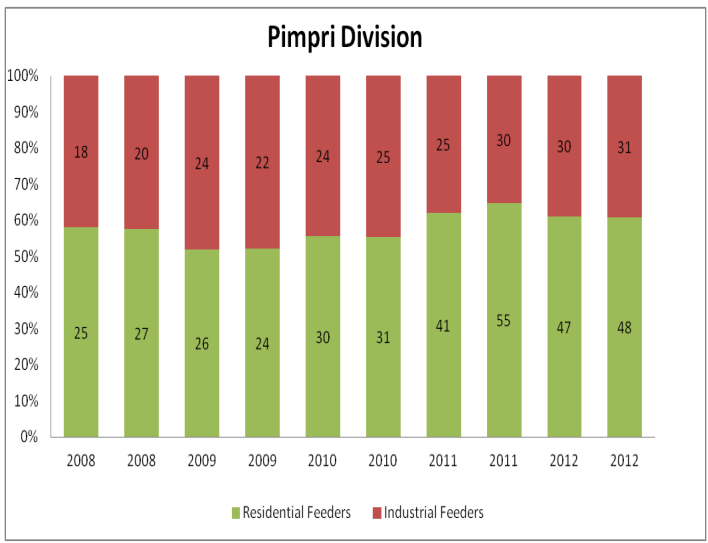
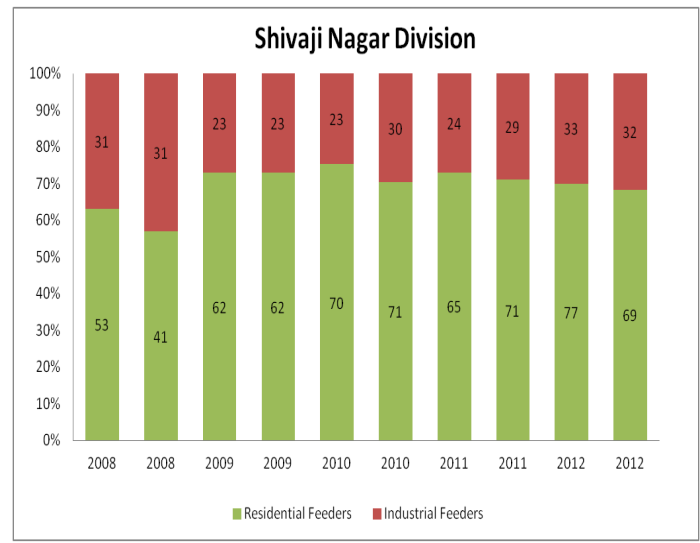

Bhosari division is dominated by industries with over 80% non - sheddable feeders, therefore classified as an industrial division. Kothrud has almost 90% sheddable feeders and hence classified as fully residential. Pimpri division shows a mix of residential and industrial activity with an equal distribution of sheddable and non sheddable feeders. Shivaji Nagar Division shows a mixed characteristic while dominated by residencies. In this division, it was observed that non sheddable supply is fed to commercial activities such as IT spaces, hotels, retail spaces, educational institutions. These activities are expected to exhibit similar temperature sensitive ED characteristics as the residencies. Therefore this division is classified as residential dominated. The classification of the feeders is shown in table 1.

| Division | Characteristics |
|---|---|
| Bhosari | Industrial dominated |
| Kothrud | Residential Dominated |
| Pimpri | Mixed Industrial and Residential |
| Shivaji Nagar | Residential Dominated |

**Table 1: Division Classification according to the feeder characteristics**

## 3.2 Statistical analysis

### 3.2.1 Estimation of apparent temperature

The daily apparent temperature was calculated using Steadman (1994) formula by adjusting the dry bulb temperature with relative humidity, following Steadman (1994). AT was calculated using the following equation -

$$AT = T_a + 0.33 * e - 0.7 * w - 4$$

Where,
$T_a$ = Daily average temperature
$rh$ = relative humidity in %,
$w$ = wind speed (m/s) at an elevation of 10 m
$e$ = water vapor pressure
$AT$ = Apparent Temperature

### 3.2.3 Extreme value analysis

The month wise trend in electricity demand was studied with the corresponding mean apparent temperature for each division. This gave a preliminary understanding of the seasonal behaviour of ED.

It was observed in the data that the number of feeders changes frequently. Therefore, to avoid the bias in the estimation of ED, the year wise anomaly in ED was plotted against AT. In order to analyse the data, we performed a regression analysis of ED with respect to the AT using year fixed effects model. The scatter plot and the fixed effects regression analysis enabled us to formulate the hypothesis for GEV analysis.

In our study, Extreme Value Theory (EVT) was used to analyse the behaviour of peak electricity demand as a function of apparent temperature. Extreme value theory (EVT) is a statistical modelling technique of extreme behaviour in observations. It is used in managing financial risks, predicting stock markets, portfolio management in the insurance industry, climate extremes, etc. It quantifies the stochastic behaviour of the processes at an extremely small scale, for example, the probability of occurrence of extreme rainfall in 100 years or the future impact of a stock market crash. These observations or tail-end of cumulative distribution function (CDF) are modelled with the family of GEV or Poisson distributions (Coles et al., 2001). The stationary model observes the stochastic behavior of independent and identically distributed random variables through time. A non–stationary model would assume that present day temperature is dependent on the previous day's temperature.

We used the block maxima approach to fit the maximas in the GEV distribution for both, stationary and non - stationary model. The advantage of this method is that it can be easily incorporated with the covariates (Coles et al.,2001). The block sizes used in our study are weekly, fortnightly and monthly. The purpose of selecting multiple blocks is to analyse the best fit of distribution over different block sizes. The analysis was performed over three blocks to reduce the arbitrariness over the choice of block size. The daily electricity demand is standardised over the blocks to reduce the bias due to changing feeders using the following equation:

$$maxima_i = (ED - mean(ED_i)) / sd(ED_i)$$

where $maxima_i$ is the block maxima for block *i*.
$ED$ is the electricity demand on day *d*.
$ED_i$ is the daily sum of electricity demand for block *i*.
$mean$ is the statistical mean.
$sd$ is the standard deviation.

We have excluded Thursdays in our analysis of Bhosari and Pimpri division because it is a non–working day for most industries, reducing the electricity demand significantly from non–sheddable feeders. Similarly, Sundays are excluded from Kothrud, Pimpri and Shivaji Nagar respectively.

The 'ismev' package in R is used to plot and analyse the GEV distribution. Block maxima, which is the standardised electricity demand, is used in the stationary model to fit into the GEV distribution. We estimate the parameters of GEV distribution using maximum likelihood method. The advantage of this method is that it can be easily incorporated with the covariates (Coles et al.,2001). In our study, apparent temperature is the covariate and fitted into the gev distribution.

Using the standard notation GEV (μ, σ, ξ) ,the non-stationary model with respect to AT is given as,

$\widehat{Z}_t \sim$ GEV ($\widehat{\mu}(t), \widehat{\sigma}(t), \widehat{\xi}(t)$ )

where $t$ is the apparent temperature. μ, σ and ξ are the location, scale and shape respectively. The non-stationary model assumes that $\mu$ and $\sigma$ are varying w.r.t $t$. It is a plausible assumption since daily temperatures are changing both in mean and variability for the various blocks.

The temperature dependence of $\mu$ and $\sigma$ is expressed as

$\widehat{\mu}(t) = \widehat{\mu}_o + \widehat{\mu}_1(t)$

$\log(\widehat{\sigma}_t) = \widehat{\sigma}_0 + \widehat{\sigma}_1 t$

### 3.2.4 Approach for Estimation

Minimization of negative log-likelihood (nllh) approach estimates the location (μ), scale (σ) and shape (ξ). The null hypothesis of no relation between the electricity demand and the apparent temperature is tested by comparing nllh for the two models

Model 1 : only μ varying linearly

Model 2 : μ varying linearly and σ varying exponentially

nllh(Model 1) = −logL(μ(t), ξ)

nllh(Model 2) = −logL(μ(t), σ(t), ξ)

The difference between negative log-likelihood values of Model 1 and Model 2 approximately follows the chi-squared distribution. The distribution was tested for 95% significance level with one degree of freedom. Therefore,

$2 * (nllh(M1) - nllh(M2)) \sim \chi^2$

$\chi^2 <= 3.84$

The GEV model enables us to test if the apparent temperature is able to explain the peak electricity demand in each of the divisions.

## 4. Results and Discussion

The study shows the results of the analysis performed between the period 2008-2012 for the four divisions. Figure 1 shows the monthly pattern of electricity demand from January to December and its variation with temperature. In the Bhosari division, it is apparent that the monthly distribution of ED is not responsive to the changes in AT. The sharp increase observed in the AT during the hot months, March and April, do not show a corresponding increase in the ED for the industrial division. On the other hand, between the months of September and December, when the apparent temperature takes a dip, the electricity demand shows a slight increase. Whereas in the mixed division of Pimpri and residential divisions of Shivaji Nagar and Kothrud, the relationship is much more conspicuous and varies more consistently with the AT line.

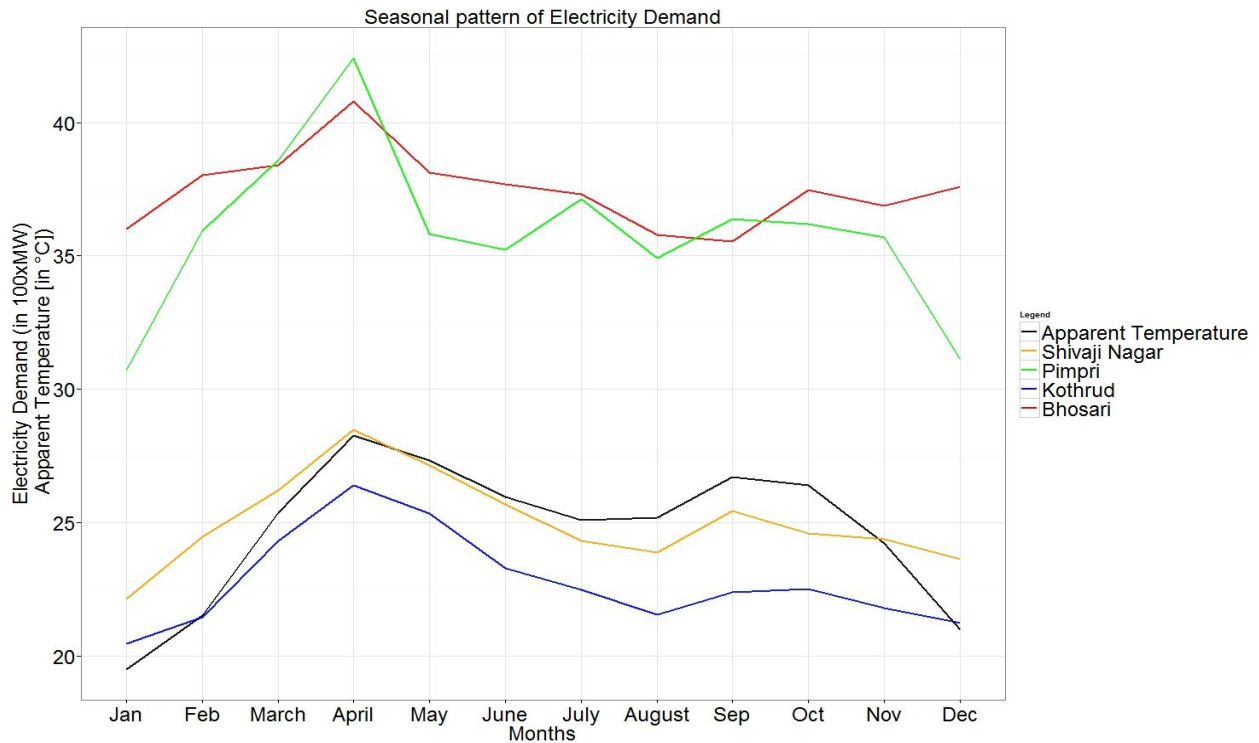

Figure 1: The graphs of Electricity Demand (in 100X MW) for each of the four divisions in different colours shows that the seasonal pattern of ED matches with the pattern of AT. The black line represents apparent temperature.

The year wise anomaly in electricity demand is plotted against the AT in figure 2. The black line represents the LOESS curve for all the years. The scatter plot for the Bhosari represents a weak correlation between anomaly in ED and the apparent temperature.

This is evident through evenly distributed points around the zero-ED line for all the years. The residential divisions viz., Kothrud and Shivaji Nagar show a steep line, indicating the anomaly in ED grows sharper with increasing temperatures. Pimpri also shows an increase in demand with the rise in temperature.

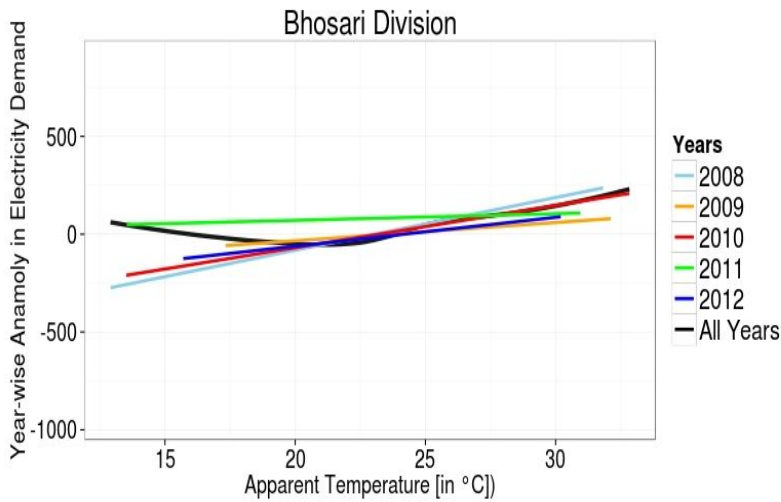
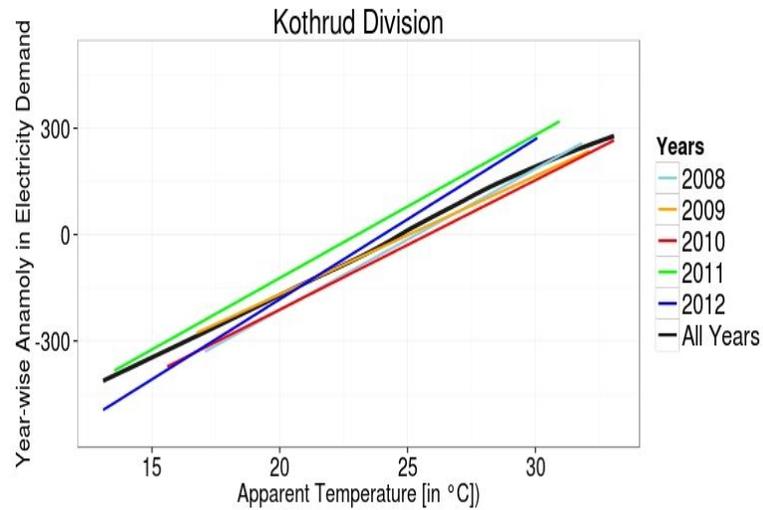
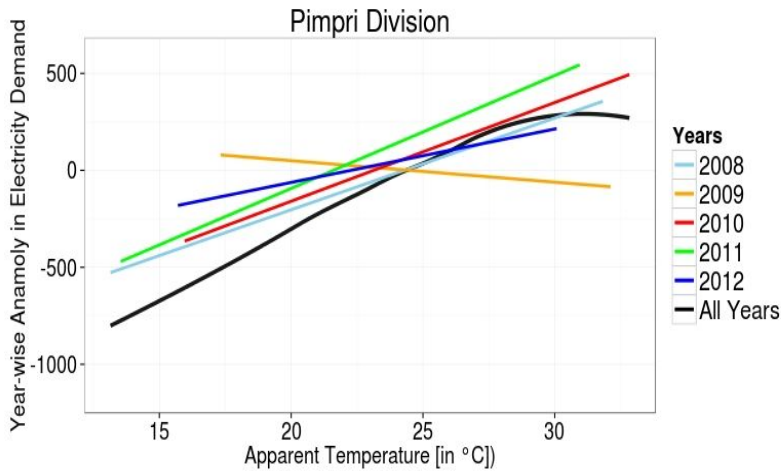
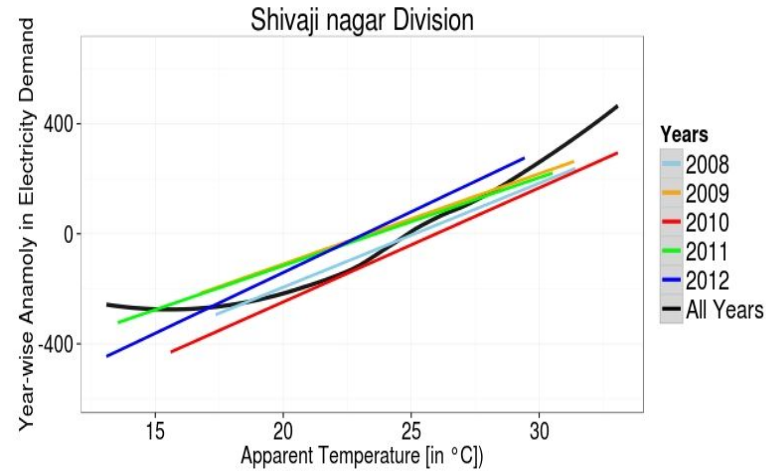

In order to explain this relationship further, a year-fixed effects model was used to control the omitted variable bias. Therefore, other factors affecting ED such as household incomes, household size, electricity price, population etc. are assumed as constant. We ran the regression model with the fixed-year effects taking 2008 as the omitted year. The results of the year – fixed effects model is shown in Table 2. All the divisions have extremely low p–values, validating the model's statistical significance.

Figure 3 shows the plot of observed and model values.

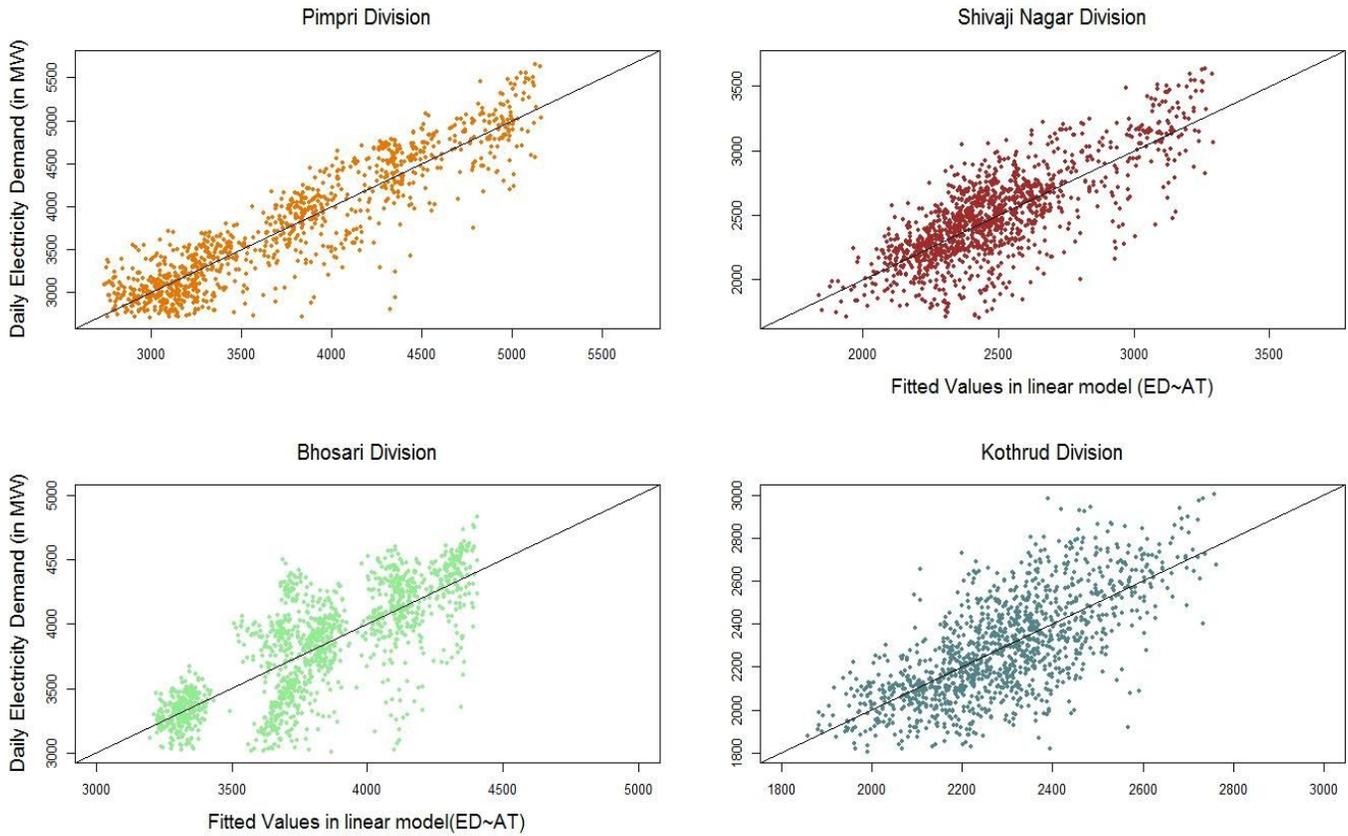

| Division | Adj R – Square | P -value | Slope |
|---|---|---|---|
| Bhosari | 0.5953 | 1.32E-12 | 15.927 |
| Kothrud | 0.4933 | 2.00E-16 | 40.813 |
| Pimpri | 0.8425 | 2.00E-16 | 45.653 |
| Shivaji Nagar | 0.5963 | 2.00E-16 | 43.438 |

**Table 2: Regression results for the fixed- year effects model.**

In Bhosari, Kothrud, Pimpri and Shivaji Nagar $1^\circ$ degree increase in AT results in additional electricity demand of 15.9, 40.8, 45.7 and 43.4 MW respectively. The daily ED in Bhosari and Pimpri is quite higher than Kothrud and Shivaji Nagar (shown in Figure 1), therefore percentage change in the Bhosari and Pimpri will be very small for per degree increase in AT as compared to the Kothrud and Shivaji Nagar. In the residential dominated Kothrud and Shivaji Nagar divisions, the mid-range values fit more in the model than the higher range values. While in the Bhosari division, the model values fit better with the observed values. Moreover, the slope of the regression line obtained by Bhosari division is far lower than that of the other divisions. These results from the model point at a way higher dependence of ED on temperature extremes in the residential areas. The fixed effect is good fit for the Pimpri division and this is partly attributed to the changing feeder characteristics of the division from industrial to residential activities during the study period.

The regression model is able to explain the behaviour of ED to the apparent temperature. However, the behaviour of peak electricity demand needs to be examined using the EVT. The block maxima approach is followed using block sizes of 7, 15 and 30 days.

### 4.1 Stationary GEV Model Analysis

The Extreme value Analysis (EVA) for stationary model of ED was performed to understand the fitting of observed values with the GEV model. Figure 4 shows the quantile plot (Q-Q plot) outlining the model values against the empirical values. The empirical values shown in the figures are the maximum of non-overlapping 15 days block for the study period 2008–2012.

The linearity of Q-Q plots denotes the goodness of fit of the distribution. In this case, the best fitting stationary model is Pimpri followed by Shivaji Nagar division.

The parametric model fit well here since graph has a linear form. The linearity of Q-Q plots defines the goodness of fit of the distribution. In our case, the best fitting stationary model is Pimpri followed by Shivaji Nagar division.

### 4.3 Non-stationary GEV model analysis

The maximum electricity demand of 15 days non-overlapping block is fitted to the GEV distribution with apparent temperature as covariate and both $\mu$ and $\sigma$ varying linearly. The residual quantile plot of the distribution is shown in Figure 5 and Table 2 shows the summary results.

The maximum electricity demand of 15 days non-overlapping block is fitted to the GEV distribution with apparent temperature as covariate and both μ and σ varying linearly. The data points are the number of block maximas available for the study period.

The likelihood ratio test values for Kothrud, Pimpri and Shivajinagar for the 15 days block is 3.60, 5.00 and 7.29. The likelihood ratio test reveals that Bhosari division is not fitted to the non-stationary model at 95% significance. It is expected due to industrially dominated activities of the division and non-dependence of demand on AT. For the Pimpri and Shivajinagar divisions, likelihood ratio test is significant at 95% significance level. At 90% significance level, Kothrud division fits with the non-stationary model. The non-stationary model is able to explain the variation in the peak electricity load better than the stationary model for the three divisions.

It is clear from the above plots that the model allowing for linear trend in μ and σ is adequate for our data. The quality of the fitted model is supported by the diagnostic plots applied to residuals. Establishing a relationship between peak load demand and the apparent temperature is useful for the future prediction of peak load as a result of global warming.

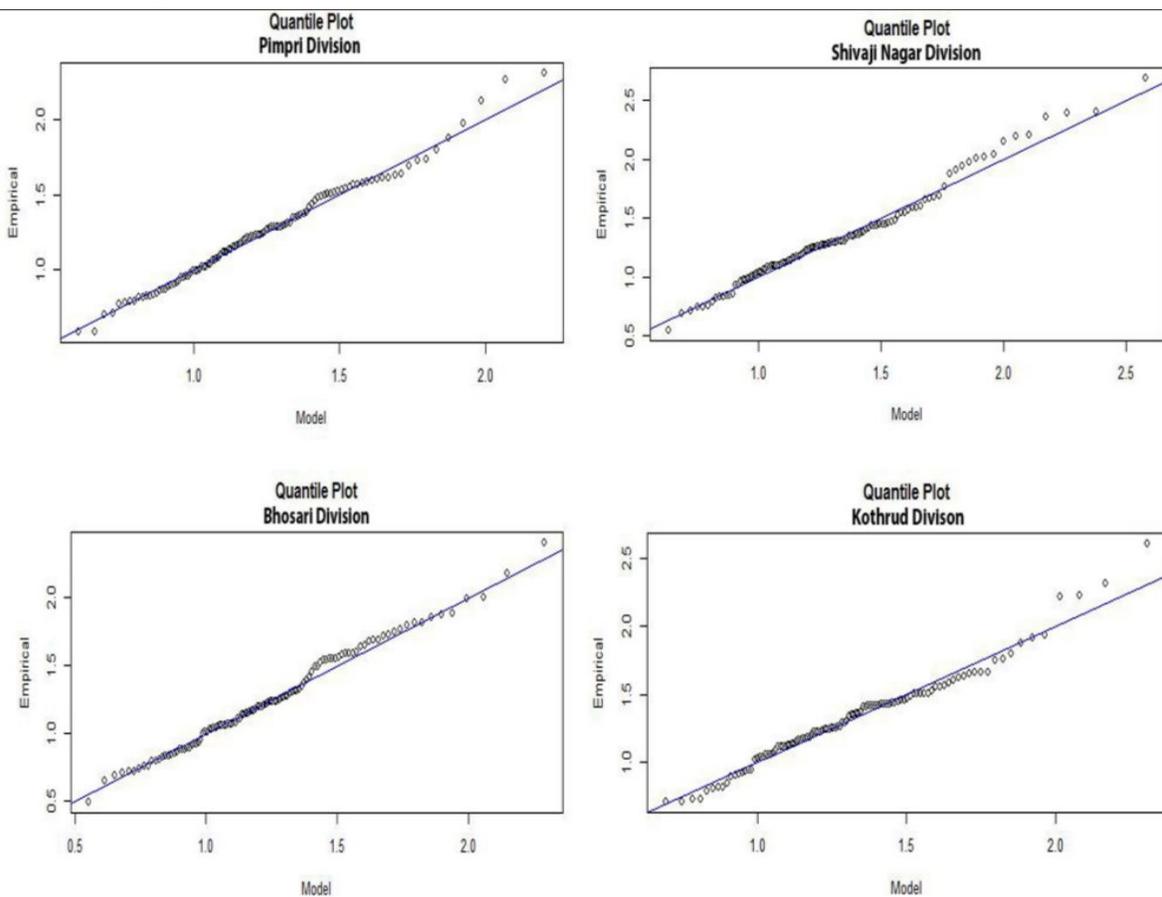

Figure 5 : Q – Q plots obtained by plotting values from the GEV stationary model against the empirical values for the four divisions.

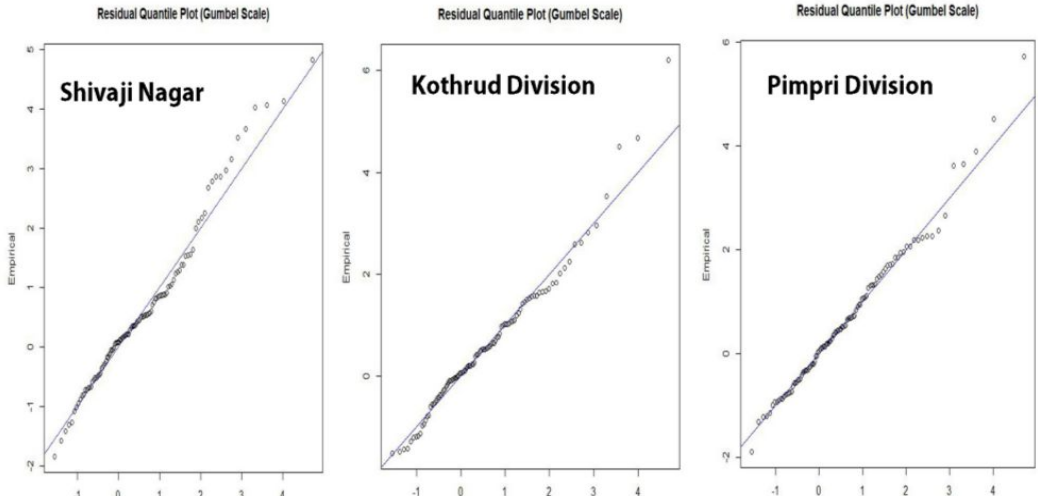

**Fig 6: Diagnostic plots to assess the goodness of fit of the non-stationary model of GEV distribution**

| Division | Data Points | | $\mu_0$ | $\mu$ | $\sigma_0$ | $\sigma$ | $\xi$ | nllh | Likelihood ratio test |
|---|---|---|---|---|---|---|---|---|---|
| **Bhosari** | 113 | MLE | 0.8587 | 0.0093 | -1.1736 | -0.0009 | -0.0981 | 42.3987 | 1.0973 |
| | | SE | 0.2239 | 0.0089 | 0.5478 | 0.0218 | 0.0738 | | |
| **Kothrud** | 108 | MLE | 0.7778 | 0.0156 | -2.1586 | 0.0358 | -0.1075 | 29.7431 | 3.6064 |
| | | SE | 0.2372 | 0.0094 | 0.7244 | 0.0282 | 0.0574 | | |
| **Pimpri** | 111 | MLE | 0.5351 | 0.0220 | -1.4974 | 0.0089 | -0.0840 | 28.5960 | 7.2977 |
| | | SE | 0.2043 | 0.0080 | 0.5252 | 0.5252 | 0.0634 | | |
| **Shivaji Nagar** | 108 | MLE | 0.5891 | 0.0219 | -1.0455 | -0.0045 | -0.0152 | 45.5706 | 5.0088 |
| | | SE | 0.2525 | 0.0098 | 0.6393 | 0.0249 | 0.0623 | | |

Table 2 : MLE is the maximum likelihood estimator while SE is the standard error. Nllh is the negative log likelihood value, likelihood ratio test is the 2*(nllh (non – stationary) - nllh (stationary)). $\mu$, $\sigma$ and $\xi$ are the location, scale and shape respectively.

## 5. Conclusion

In this paper, we addressed the problem of the impact of the change in electricity demand due to rising temperatures. For the developing country like India located in sub-tropical and humid climate, this study quantified the change in electricity demand attributed to the change in temperatures. The key contribution of this paper is to present the disaggregated analysis of industrial and residential activities. We quantified the impact of rising temperature on the electricity demand in industries and residencies. The study captures the additional peak load on power infrastructure due to global warming.

The role of climate in influencing peak load conditions is significant, and it is shown in the analysis of the industrial and residential areas of the four divisions of Pune. The classification of divisions by distinguishing the feeders by sheddable and non–sheddable characteristics was quintessential to obtain our results. It is quite evident from the regression and the fixed-year models that the electricity demand to residential divisions exhibits more temperature sensitivity during summers than industrial areas. The study predicts that there will be a significant rise of peak electricity load as a consequence of the increase in temperatures, by keeping the other factors as constant. The fitness of the stationary and non-stationary GEV models with the observed data, suggests that the Extreme Value Theory can be used to predict future peak electricity demand combined with extreme temperature events, which are expected to recur with greater frequency in a global warming scenario. This analysis brings about a need to enhance existing electricity distribution infrastructure in a rapidly urbanising Pune in the context of global climate change.

As part of our future work, we plan to quantify the additional peak load on electricity infrastructure as a consequence of various RCP scenarios under IPCC climate change predictions. Since electricity load data is now increasingly available in the public domain, we will also extend and apply our technique to other regions. Finally, we will explore the better extreme value models to complement our approach described in this paper.